\documentclass[
superscriptaddress,
twocolumn,
nofootinbib,
 amsmath,amssymb,
 aps,
pra,
showpacs
]{revtex4-1}
\usepackage{comment}
\usepackage{hyperref}
\usepackage{graphicx}
\usepackage{dcolumn}
\usepackage{bm}
\usepackage{simplewick}
\usepackage{color}
\usepackage[utf8]{inputenc}
\usepackage{dirtytalk}
\usepackage{placeins}

\newcommand{\Z}{\mathcal{Z}}
\newcommand{\I}{\left<I\right>}
\newcommand{\A}{\mathcal{A}}
 \newcommand{\bbar}{{\mkern0.75mu\mathchar '26\mkern -9.75mu b}}
 \newcommand{\be}{\begin{eqnarray}} 
  \newcommand{\ee}{\end{eqnarray}}
 \newcommand{\nn}{\nonumber}
 \newcommand{\F}{\int_0^\beta d\tau \int d^2\vec{x}\left[\psi _\sigma ^*\left(\partial_\tau-\frac{\nabla^2}{2}-\mu \right)\psi _\sigma +c\psi _\uparrow ^*\psi _\downarrow ^*\psi _\downarrow \psi _\uparrow \right]}
  
  \newcommand{\Fbo}{\int_0^\beta d\tau \int d^2\vec{x}\psi _\uparrow ^\dag\psi _\downarrow ^\dag\psi _\downarrow \psi _\uparrow}
  \newcommand{\Fl}{\int_0^\beta d\tau \int d^2\vec{x}\left[\psi _\sigma ^*\left(\partial_\tau-\frac{\nabla^2}{2}-\tilde{\mu} \right)\psi _\sigma +c^\lambda\psi _\uparrow ^*\psi _\downarrow ^*\psi _\downarrow \psi _\uparrow \right]}
  \newcommand{\FT}{\int_0^{\tilde{\beta}} d\tilde{\tau} \int d^2\tilde{\vec{x}}\left[\psi _\sigma ^{*\lambda}\left(\partial_{\tilde{\tau}}-\frac{\tilde{\nabla}^2}{2}-\mu \right)\psi _\sigma^\lambda +c^\lambda\psi _\uparrow ^{*\lambda}\psi _\downarrow ^{*\lambda}\psi _\downarrow ^\lambda\psi _\uparrow^\lambda \right]}

 \newcommand{\m}{\left[\prod_{\sigma}d\psi_\sigma^*d\psi_\sigma\right]}
 \newcommand{\ml}{\left[\prod_{\sigma} d\psi_\sigma^{*\lambda}d\psi_\sigma^\lambda\right]}
\newcommand{\mb}{\frac{\left[d\phi^*d\phi\right]}{N}}
\newcommand{\D}{\mathcal{D}}

\newcommand{\fkD}{\frac{d^2k}{(2\pi)^2}}
\newcommand{\fkDd}{\frac{dk^2}{(2\pi)^2}}
\newcommand{\fw}{\frac{d\omega}{(2\pi)^2}}

\usepackage{graphicx}
\usepackage{placeins}
\usepackage{dirtytalk}
\usepackage{simplewick}
\usepackage{pgf,tikz}
\usepackage{mathrsfs}
\usetikzlibrary{arrows}
\usetikzlibrary{calc,decorations.markings}
\usetikzlibrary{arrows.meta,shapes.misc,decorations.pathmorphing,calc,bending}

\tikzset{
  branch point/.style={cross out,draw=black,fill=none,minimum size=2*(#1-\pgflinewidth),inner sep=0pt,outer sep=0pt}, 
  branch point/.default=5
}
\tikzset{
  branch cut/.style={
    decorate,decoration=snake,
    to path={
      (\tikztostart) -- (\tikztotarget) \tikztonodes
    },
    execute at begin to={{
      \coordinate (A) at ($(\tikztostart)!.8!-10:(\tikztotarget)$);
      \coordinate (B) at ($(\tikztostart)!.8!10:(\tikztotarget)$);
      \coordinate (AB/3) at ($(A)!1/3!(B)$);
      \coordinate (2AB/3) at ($(A)!2/3!(B)$);
      \coordinate (C) at ($(AB/3)!2/(3*sqrt(3))!-90:(B)$);
      \coordinate (D) at ($(2AB/3)!4/(3*sqrt(3))!-90:(B)$);
      \draw[thick,green!60!black,-{Stealth[]}] (A) .. controls (C) and (D) .. (B) node[scale=.8,pos=.9,above left] {$\times (-1)$};
    }}
  }
}

\begin{document}

\preprint{APS/123-QED}

\title{Virial expansion for the Tan contact and Beth-Uhlenbeck formula from two dimensional $\mathbf{SO(2,1)}$ anomalies}

\author{W. Daza}
\email{wsdazaromero@uh.edu}
\affiliation{Physics Department, University of Houston, Houston, Texas 77004, USA\\}

\author{J. E. Drut}
\email{drut@email.unc.edu}
\affiliation{Department of Physics and Astronomy, University of North Carolina, Chapel Hill, North Carolina 27599, USA\\}

\author{C. Lin}%
 \email{cllin@uh.edu}
 \affiliation{Physics Department, University of Houston, Houston, Texas 77004, USA\\}
 
 \author{C. Ord\'o\~nez}%
 \email{ordonez@uh.edu}
\affiliation{Physics Department, University of Houston, Houston, Texas 77004, USA\\}






\date{\today}

\begin{abstract}
The relationship between 2D $SO(2,1)$ conformal anomalies in nonrelativistic systems and the virial expansion is explored using recently developed path-integral methods. In the process, the Beth-Uhlenbeck formula for the shift of the second virial coefficient $\delta b_2$ is obtained, as well as a virial expansion for the Tan contact.  A possible extension of these techniques for higher orders in the virial expansion is discussed.
\end{abstract}

\maketitle

\section{\label{sec:level1} Introduction}

The virial expansion has been widely used in the study of strongly correlated systems and in other contexts, as it captures the impact of few-body physics on the high-temperature thermodynamics of many-body systems. The expansion has recently been used in the study of ultracold atomic Fermi gases \cite{Liu201337,Chafin:2013zca}, where the realization of 2D systems has now been achieved by multiple groups around the world (see e.g.~\cite{RanderiaPairingFlatLand, PieriDanceInDisk}). 
While the most common form of the expansion is that of the pressure equation of state, of particular interest is the virial expansion of the Tan contact \cite{Tan20082971}, as the latter determines all short-range correlations in systems with contact interactions. The calculation of virial coefficients, however, is a challenging problem: in its most straightforward form, computing the $n$-th order requires solving the $m$-body problem for all $m \leq n$. Thus, a number of different approaches have been proposed to calculate the virial coefficients, all of which aim at producing a reliable and efficient computational scheme \cite{Kaplan:2011br,Leyronas} that bypasses finding such full solution. 

In 2D, the existence of a scaling anomaly provides an appealing conceptual framework to establish relationships between different relevant aspects of these calculations, as well as hints for a possible systematic procedure for higher-order coefficients. A signal of the connection between the virial expansion and 2D anomalies is already present in the celebrated Beth-Uhlenbeck (BU) formula for $\delta b_2$, the shift from the free value of the second virial coefficient, which in the case of 2D attractive contact interactions of nonrelativistic Fermi particles becomes \citep{Chafin:2013zca}
\be \label{uno}
\delta b_2=e^{\beta E_b}-2\int \frac{dk}{k}\frac{e^{-2\beta\epsilon_{k}}}{\pi^2+\ln^2(\frac{k^2}{E_b})}.
\ee
%

Here $E_b$ is the  magnitude of the single bound state energy allowed by this system.  The first term comes precisely from the presence of this bound state, and the integral term comes from the scattering sector, once the phase shift for the s channel has been properly accounted for.  This system possesses an SO(2,1) symmetry~\cite{PhysRevA.55.R853,jackiw1991mab}, which includes scaling symmetry.  If the symmetry is respected at the quantum level, the bound state term in (\ref{uno}) would not be included (the existence of a finite energy $E_b$ would provide a scale in the system, hence breaking the classical scaling symmetry). The scattering term in the original BU formula contains the derivate of the phase shift with respect to the momentum or energy; if the scaling symmetry is preserved, this term would be zero. Therefore, this heuristic argument seems to signal a direct relationship between 2D anomalies and $\delta b_2$.  

In this paper, we show that $\delta b_2$ is indeed produced entirely by the anomaly. We use a path-integral approach inspired by the work of \citep{Ordonez:2015vaa,LinCarlos,pathintegral1,pathintegral2,pathintegral3,pathintegral4,Lin:2016shf}. In the process, we will describe the virial expansion of the Tan contact (which in 2D is interpreted as the anomaly \cite{Hofmann:2012np}), as well as a procedure to compute $\delta b_n$, $n\geq 2$, using the Hubbard-Stratonovich  (HS) representation of the partition function. 

The rest of the paper is organized as follows:  In section II we will derive the anomaly, showing the identification with the Tan contact; in this section we will also relate the anomaly with the virial expansion and will derive the main general formula for $\delta b_n$; the explicit calculation for $\delta b_2$ and its connection with the BU formula will be shown. Section III will sketch the procedure to calculate $\delta b_n$ and the first results on $\delta b_3$ will be discussed. Section IV will contain conclusions and comments. We would like to emphasize that our goal in this paper is to lay out the framework more than to engage in applications, although we will naturally connect with other approaches to assess similarities and differences. We hope that our approach will offer insight into the questions addressed here. 



\section{Anomaly, Tan Contact, virial expansion and derivation of the Beth-Uhlenbeck formula}
\subsection{Structural Aspects}

\noindent The partition function for a 2D dilute gas of nonrelativistic spin $1/2$ Fermions is\footnote{In this paper $\hbar=k_B=m=1.$}
\be
\label{pI}
\mathcal{Z}=\text{tr}\left[ e^{-\beta\left(H-\mu N\right)}\right]=\int\m e^{-S_E},
\ee
where
\be
S_E=\F\nn.
\ee
The Fermion fields have antiperiodicity  $\beta.$ The index $\sigma$ is summed over $\uparrow, \downarrow$ values in the Euclidean action term. Following \cite{Chafin:2013zca}, the dimensionless coupling constant $c$ will be selected to incorporate the nonperturbative physics connected with the existence of a bound state. For the attractive case, the Lippmann-Schwinger equation gives the pole of the scattering matrix $T$ describing the bound state energy $E_b$ of the 2-body problem \citep{Phillips:1997xu,Weinberg:1995mt}:
\be \label{Ls}
T(p',p,E)&=&V(p',p)+\\
&\quad& +\int\frac{d^2k}{(2\pi)^2}V(p',k)\frac{1}{E-\vec{k}^2+i\epsilon}T(k,p,E).\nn
\ee
In momentum space, the Dirac delta potential is $V(p',p)=c$. From the previous equation one gets
$$\frac{1}{T(E)}=\frac{1}{c}-\int\frac{d^2k}{(2\pi)^2}\frac{1}{E-\vec{k}^2+i\epsilon}.$$
At the bound state, $1/T(-E_b)=0$, $E_b>0$, such that
\be 
\label{cexp}
\frac{1}{c}=\frac{1}{c\left(\frac{E_b}{\Lambda^2}\right)}&=&\frac{1}{(2\pi)^2}\int^{\Lambda\to\infty}\frac{d^2\vec{k}}{-E_b-\vec{k}^2+i\epsilon}\\
&=&\frac{1}{4\pi}\ln{\left(\frac{E_b}{\Lambda^2}\right)}+\text{Finite constant}.\nn
\ee
The expression for $\frac{1}{c}$ is singular, and we choose to regularize it with a large cutoff $\Lambda$. This infinity will be used to cancel a divergence that will arise in the calculation of the effects of the interaction in the path integral of Eq.~(\ref{pI}).

As is well known, the action $S_E$ has a classical invariance under the following scaling transformations [part of $SO(2,1)$ invariance]:
\be 
\tau&\to&\tilde{\tau}=\lambda^2\tau,\nn\\ \vec{x}&\to&\tilde{\vec{x}}=\lambda\vec{x},\\ \psi(\tau,\vec{x})&\to&\psi^\lambda(\tilde{\tau},\tilde{\vec{x}})=\lambda^{-1}\psi(\tau,\vec{x}).\nn
\ee

Using dimensional analysis, the following equation was derived in Ref.~\cite{Lin:2016shf} (see appendix A):
\begin{equation}
\label{6equation}
2 \mathcal E-DP=-2\sum\limits_k E_k \frac{\partial P}{\partial E_k},
\end{equation}
where $\mathcal E=\text{energy density} = \frac{\langle H \rangle}{A}$, $A=\text{2}D\text{ volume}$, $P$ is the pressure, and $D$ the dimensionality of space ($D$=2 in this paper). The $\{E_k\}$ are a set of energy parameters that may include bound state energies as well as those formed from dimensionful couplings constants in $S_E$ \cite{Lin:2016shf}.
In our case, there is no dimensionful coupling constant ($c$ is dimensionless) and there is only one bound state energy $-E_b$, $E_b>0$ (we will use $E_b$ in  Eq. \eqref{6equation} henceforth), such that
\begin{equation}
2 \mathcal E-2P=-2 E_b \frac{\partial P}{\partial E_b}.
\end{equation}
Now, $\frac{\partial}{\partial E_b}=\frac{\partial c}{\partial E_b}\frac{\partial}{\partial c}$, and from Eq. \eqref{cexp} $\frac{\partial c}{\partial E_b}=-\frac{c^2}{4\pi E_b}$. Therefore
\begin{equation}
2 \mathcal E-2P=\frac{c^2}{2\pi}\frac{\partial P}{\partial c}.
\end{equation}
In the thermodynamic limit ($A \rightarrow \infty$, $\Omega \rightarrow -PA$),
$\beta P A=\ln \mathcal Z$, hence\footnote{$\langle \hat{\theta} \rangle=\frac{\text{tr}\,\left(e^{-\beta\left(H-\mu N\right)}\hat{\theta}\right)}{\text{tr}\,e^{-\beta\left(H-\mu N\right)}}.$}
\begin{equation}\label{9equation}
\frac{\partial P}{\partial c}=\frac{1}{\beta A \mathcal Z}\frac{\partial \mathcal Z}{\partial c}=\frac{-4\pi}{Ac^2}\langle I \rangle,
\end{equation}
where
\begin{equation}
I=\frac{c^2}{4\pi}\int  d^2\vec{x} \, \psi^\dagger_\uparrow \psi^\dagger_\downarrow  \psi_\downarrow \psi_\uparrow,
\end{equation}
which is Tan's contact.
Here, we used the fact that, in equilibrium, $\langle \psi^\dagger_\uparrow \psi^\dagger_\downarrow  \psi_\downarrow \psi_\uparrow \rangle$ is $\tau-$independent to derive Eq. \eqref{9equation}.
The scaling anomaly \cite{Hofmann:2012np,PhysRevD.46.5474} is therefore
\begin{equation} \label{equation11}
\mathcal{A}=2P-2\mathcal E=\frac{2}{A}\langle I \rangle.
\end{equation}
In Appendix B we also prove this result using the ideas and techniques of Ref.~\cite{Ordonez:2015vaa}.

\subsection{Virial expansion for the anomaly}

From Eq. \eqref{9equation}, Tan's contact can be written as
\be 
\I=-\frac{1}{4\pi\beta\Z}c^2\frac{\partial}{\partial c}\Z.
\ee
Writing $\Omega=\Omega^{free}+\delta\Omega$, where $\delta\Omega$ is the contribution from interactions, $\Z$ can be expressed as
\be \label{19}
\Z&=&e^{-\beta\Omega^{free}}e^{-\beta\delta\Omega}\nn\\
&=&\Z^{free}\Z_I.
\ee
Using   (\ref{cexp}), $\frac{\partial}{\partial c}=\frac{\partial E_b}{\partial c}\frac{\partial}{\partial E_b}=-4\pi E_bc^{-2}\frac{\partial}{\partial E_b}$ gives
 \be 
 \frac{\partial \Z}{\partial c}=4\pi\beta c^{-2}\Z E_b \frac{\partial}{\partial E_b}\delta\Omega,
 \ee
 and hence
 \be 
 \label{tan}
 \I=-E_b\frac{\partial}{\partial E_b}\delta\Omega.
 \ee
 The anomaly becomes
  \be \label{23}
 \A=-2E_b\frac{\partial}{\partial E_b}\left(\frac{\delta\Omega}{A}\right).
 \ee
 Defining the virial expansion by\footnote{We use two species of Fermions $(\uparrow\, , \, \downarrow).$} \cite{Parish}
\begin{align}\label{Omega}
\delta\Omega &\equiv  \sum_{n\geq 2}z^n\delta \Omega_n\\
&=-\frac{1}{\beta^2\pi}\sum_{n\geq 2}z^n\delta b_n,\tag{17a}
\end{align} 
where $z=e^{\beta\mu}$ is the fugacity\footnote{One should recall that when $T$ is sufficently large the Fermion gas will behave as a classical gas. Therefore the chemical potential $\mu$ will become negative as it is for classical gases \cite{chemical}. Consider the product $\beta\mu$ in this limit, for 2D:
\begin{eqnarray}
\beta\mu=\beta\left(-\frac{1}{\beta}\ln\left[F\frac{T}{\rho}\right]\right),\nn
\end{eqnarray}
where $\rho $ is the density of the system and $$F=g(mk/2\pi\hbar^2)$$ whith $k$ the Boltzman constant and $g$ the degeneracy of the particles, $g=2s+1$, for particles of spin $s$.  Therefore, in the limit $T\to \infty$ (notice that we also have $\rho \to 0$) the product $\beta\mu\to -\infty$. If we now define the fugacity as $z=e^{\beta\mu}$ we see the at order zero in the fugacity , i.e.,
\begin{eqnarray}
\lim_{T\to \infty}z=0.\nn
\end{eqnarray}}.
Equation (\ref{tan}) then becomes
\be \label{equation18}
\I=\frac{1}{\beta^2\pi}E_b\sum_{n\geq 2}z^n\frac{\partial}{\partial E_b}\delta b_n,
\ee
and Eq.~(\ref{23}) becomes
\be 
\A=\frac{2E_b}{\pi\beta^2}\sum_{n\geq 2}z^n\frac{\partial}{\partial E_b}\delta \bbar_n,
\ee
where $\delta \bbar_n= {\delta b_n}/{A}$.

The anomaly $\A$ can also be formally computed using the knowledge of $\Omega=\Omega^\text{free}+\delta\Omega$  \footnote{Use $\langle H \rangle=\Omega-T\partial\Omega/\partial T-\mu\partial\Omega/\partial \mu,$ $P=-\Omega/A $ (infinite $A$ limit) to compute $2P-2\mathcal{E}$. We also used the fact that all $\mu$ dependence in $\delta \Omega$ is captured by the fugacity $z$; this is best seen from the standard definition of the virial expansion, $Z=\text{tr}\,\left(e^{-\beta(H-\mu N)}\right)=\sum\limits_N z^N \text{tr}_N\,\left(e^{-\beta H}\right)$, where, by definition, all the $\mu$ dependence is therefore contained in $z$.} and the above virial expansion
\be \label{26}
\A&=&-\frac{2}{\pi\beta^2}\sum_{n\geq 2}z^n T\frac{\partial}{\partial T}\delta \bbar_n\nn\\
&=&\frac{2}{\pi\beta}\sum_{n\geq 2}z^n \frac{\partial}{\partial \beta}\delta \bbar_n.
\ee
Using Eqs.~(\ref{23}), (\ref{Omega}) and (\ref{26}) we get
\be \label{27}
\frac{1}{\pi\beta}\frac{\partial}{\partial \beta}\delta\bbar_n=-E_b\frac{\partial}{\partial E_b}\left(\frac{\delta\Omega_n}{A}\right),
\ee
or
\be \label{28}
\delta b_n=-\pi E_b\int^\beta d\beta '\beta '\frac{\partial}{\partial E_b}\left(\delta\Omega_n\right).
\ee
Equation (\ref{28}) is one of the main results in this work.
%
Notice it is defined up to an integration constant, more of which will be said below. One has then to compute $\delta\Omega_n$ in order to find $\delta b_n$. The most efficient way to do this is by means of the Hubbard-Stratonovich representation of the partition function \footnote{$N$ is the normalization constant obtained when path-integrating over $\phi^*,\phi$ in trading the quadratic Fermionic interaction for (at most) quadratic terms in the action. See \url{http://www.weizmann.ac.il/condmat/oreg/sites/condmat.oreg/files/uploads/tutorial11.pdf}.}
\be 
\label{hs}
\mathcal{Z}&=&\int\mb e ^{\left[\text{tr}\ln G^{-1}+\int d\tau \int d^2\vec{x}\frac{|\phi|^2}{c}\right]}\\
&=&\int\mb e ^{-S_\text{eff}(\phi^*,\phi,\mu)},\label{31}
\ee
with
\be \label{inverses}
G^{-1}&=&\left(\begin{matrix}
\partial_\tau-\frac{\nabla^2}{2}-\mu&\phi\\
\phi^*&\partial_\tau +\frac{\nabla^2}{2}+\mu
\end{matrix}
\right)\nn\\
&\equiv & \left(\begin{matrix}
G_1^{-1}&\phi\\
\phi^*&G_2^{-1}
\end{matrix} \right).
\ee

\subsection{Calculation of ${\delta b_2}$: Beth-Uhlenbeck formula}

We will illustrate this for the case $n=2$, for which we need to keep only up to the quadratic terms in $S_\text{eff}.$ At this point we can follow Ref.~\cite{Chafin:2013zca} and write $\Z$ as in Eq.  (\ref{19})
\be 
\Z&=&e^{-\beta\Omega^\text{free}}e^{-\beta\delta\Omega}\nn,
\ee
\be \label{32}
\delta\Omega&=& \frac{1}{2\pi i}\int_{-\infty}^{\infty}d\omega\int\frac{d^2k}{(2\pi)^2}\text{Disc}\left\{\ln\mathcal{D}^{-1}(\omega+i\epsilon,k)\right\}f_\text{BE}(\omega)\nn\\
&=& z^2\delta\Omega_2+z^3\delta\Omega_3+O(z^4),
\ee
where $f_\text{BE}(\omega)=(e^{\beta\omega}-1)^{-1}$ is the Bose-Einstein distribution function for the frequency $\omega$. To extract the $z^2$ contribution from Eq. (\ref{32}) we will use the zeroth-order ($z^0$) version of $D^{-1}(\omega+i\epsilon,k)$ given by \footnote{$\delta \Omega$ in   (\ref{32}) contains an infinite number of powers $z^n$ ($n\geq 0$) but to obtain $\delta \Omega _2$ we only retain the zeroth-order part of $D^{-1}(\omega, k).$ To find the complete contributions for higher $z^{n}, n\geq 3,$ one also has to consider the contributions from the higher powers in the effective action (see below).},

\be \label{33}
D^{-1}(\omega+i\epsilon,k)=\frac{1}{4\pi}\ln\left(-\frac{\omega+i\epsilon+2\mu-\frac{\varepsilon_k}{2}}{E_b}\right),
\ee
with $\varepsilon_k=\frac{k^2}{2}.$

We recognize several regions for the $\omega$ integration as seen in Table I. The discontinuities here been computed by studying the branch cuts of the complex logarithmic function. Since some of the sub-integrals have $E_b-$dependent limits we have to be careful with taking the $E_b$ derivatives. We obtain 
(See Appendix C for details)

\be  \label{deriv}
\frac{\partial\delta \Omega_2}{\partial E_b}=-\frac{e^{\beta E_b}}{\pi\beta} -\frac{2}{\pi\beta}\int_0^\infty d\tilde{k} \tilde{k}\frac{e^{-\beta\tilde{k}^2}}{E_b\left(\pi^2+\ln^2(\frac{\tilde{k}^2}{E_b})\right)}.\quad
\ee

\begin{figure}[t]
		\begin{center}
		\includegraphics[scale=0.7]{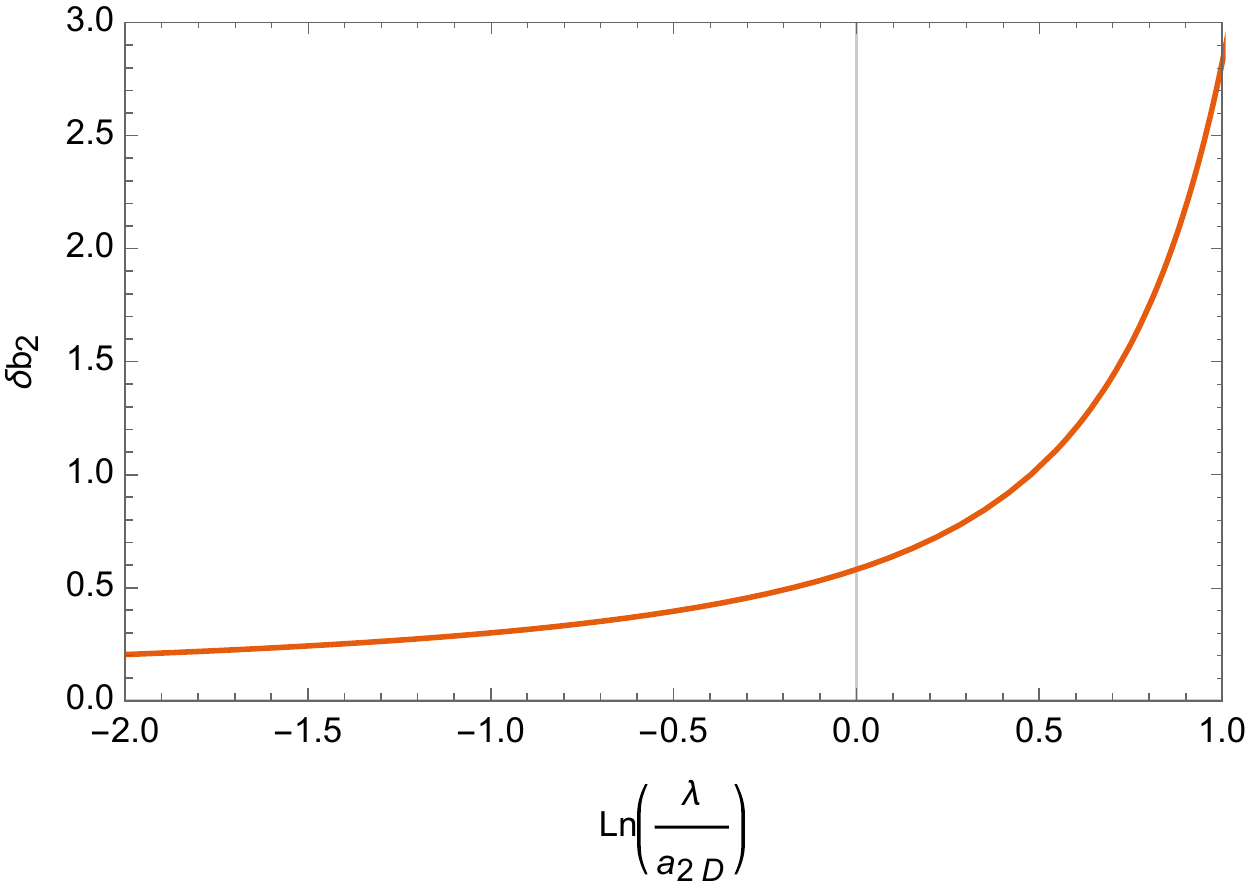}
		\caption{Shift in the second virial coefficient $\delta b_2$ as a function of the physical coupling $\ln \lambda/a_\text{2D}$, where $\lambda = \sqrt{2\pi \beta}$ is the thermal wavelength.}
		\label{secondvc}
	\end{center}
	\end{figure}

\FloatBarrier
\begin{table}[h]
\caption{$\omega$ range. Here $f=\arctan{\left({\pi}/{\ln{\left(\frac{\omega+2\mu-{(1/2)\varepsilon_k}}{E_b}\right)}}\right)}.$}
\centering
\begin{tabular}{|c|c|c|}

\hline \hline
$\omega$&$\text{Disc}\left\{\ln\mathcal{D}^{-1}(\omega+i\epsilon,k)\right\}$ \\
\hline
$(-\infty,\varepsilon_k/2-2\mu-E_b)$ &0 \\
$(\varepsilon_k/2-2\mu-E_b,\varepsilon_k/2-2\mu)$&$-2\pi i$\\
$(\varepsilon_k/2-2\mu,\varepsilon_k/2-2\mu+E_b)$&$-2\pi i-2if $\\
$(\varepsilon_k/2-2\mu+E_b,\infty)$&$-2if$\\
\hline
\end{tabular}
\end{table}
\FloatBarrier
\noindent Finally, using Eq. (\ref{28}), 

\be \label{135}
\delta b_2&=&-\pi E_b\int^\beta d\beta '\beta '\frac{\partial \delta\Omega_2}{\partial E_b}\nn\\
&=& e^{\beta E_b}-\int^\infty_0\frac{dy}{y}\frac{2e^{-\beta E_b y^2}}{\pi^2+4\ln^2 y}
.
\ee
This is the well known Beth-Uhlenbeck formula for $\delta b_2$ (rescaled verison of Eq. (\ref{uno})). The overall integration constant in   \eqref{135} is chosen to be zero so that in the limit $\delta\Omega_2\to 0$ we recover the free case. We now compare our procedure with that of \cite{Chafin:2013zca}. They obtain $\delta b_2$ by \textit{explicitly} computing $\delta \Omega _2$, and then reading off the coefficient of the $z^2$ term. We do \textbf{not} have to obtain an explicit expression for $\delta \Omega _2,$ which contains several terms in integral form with complicated integrands. The authors of \cite{Chafin:2013zca} resort to first computing $\partial \delta \Omega_2/\partial \mu$ in order to get a more manageable expression, and then perform an integral over $\mu$ to obtain $\delta \Omega_2.$ In  our case, while the original integral expressions in $\delta\Omega_2$ are complicated, $\frac{\partial\Omega_2}{\partial E_b}$ is easily calculated and given by Eq. (\ref{deriv}). We then perform a simple integration over $\beta$ to get $\delta b_2.$ We hope that similar simplifications will occur for higher $\delta b_n (n\geq 3)$. One can plot the second virial coefficient as seen in Fig. \ref{secondvc}. \footnote{We are plotting $\delta b_2$ vs $\ln\left(\frac{\lambda}{a_{2D}}\right)$, $a_{2D}=$ 2D scattering length, to compare with ref. \cite{Parish}. Here $\ln\left(\beta E_b \right)=2\ln\left(\frac{\lambda}{a_{2D}}\right)-\ln(2\pi)$. }
This result agrees with those in the literature, in particular with \cite{Parish}. The corresponding term in the virial expansion Eq.\eqref{equation18}, $I\equiv\sum\limits_{n \geq 2}z^n I_n$ is\footnote{The n-th term is $\frac{E_b}{\beta^2 \pi}\frac{\partial}{\partial E_b}\delta b_n=-\frac{E_b^2}{\beta^2}\int^\beta d\beta' \, \beta' \frac{\partial^2}{\partial E_b^2}\left(\delta \Omega_n\right).$}

\begin{equation}
I_2=\frac{E_b}{\beta^2 \pi }\frac{\partial \delta b_2}{\partial E_b}=
\left(\frac{E_b}{\beta \pi} \right) e^{\beta E_b}\left[1+2\int^\infty_0 dy\, \frac{y \,e^{-\beta E_b(y^2+1)}}{\pi^2+4\ln^2 y} \right],
\end{equation}

\noindent which agrees with ref. \cite{PhysRevLett.115.115301}, after the identification $I_2=\frac{1}{2\beta^2 \pi}c_2$ is made.
\section{Extension for ${\delta b_n}$, ${n \geq 3}$}

\subsection{General Framework}

The emphasis in this paper is on the close connection between 2D anomalies and the virial expansion for the Tan contact. Eqs.~\eqref{equation11}-\eqref{equation18} accomplish this, and in particular, Eq. \eqref{135} reflects this relationship for $\delta b_2$. As a bonus, this formulation naturally suggests a procedure to compute $\delta b_n, n\geq 3.$ In this section we will give a sketch of the procedure and will report on partial results for $\delta b_3$. While complete analytical and numerical results will be reported elsewhere, we show here that even though the complexity of the details increases, the methodology itself is a direct extension of the calculations for $\delta b_2$.

We begin by writing Eq. (\ref{31}) as
\be 
\Z=\Z_\text{free}\int\frac{[d\phi^* d\phi]}{N}e^{-(S_2+\delta S)},
\ee
where $S_2$ is the quadratic piece of $S_\text{eff}$ that gives the entire contribution to $\delta \Omega_2$,\footnote{Here we use  ``covariant notation'', i.e., $x=(\tau,\vec{x})$, etc.} namely
\be 
S_2 = \int dx dy \, \phi^*(y)\Delta^{-1}(y-x)\phi(x),
\ee
with
\be
\Delta^{-1}(y-x) \equiv -\frac{1}{c}\delta(x-y)+G_1(x-y)G_2(y-x)\quad
\ee
and $\delta S$ contains an infinite number of nonlocal terms with even powers ($2n$) of the fields, $n\geq 2.$\footnote{$G_1$ and $G_2$ were defined in   (\ref{inverses}).} One can then use the standard expansion of the exponential and the Wick theorem to calculate $\Z$ \cite{Weinberg:1995mt}, where the contraction between $\phi^*(y)$ and $\phi(x)$ is 
\be 
 \contraction{}{\phi}{^*(y)}{\phi}\phi^*(y)\phi(x) \equiv \Delta(y-x).
\ee
The first term in $\delta S$ is 
\be \label{equation33}
S_{4}&=&\frac{1}{2}\int \prod_{i=1}^4dx_i\phi^*(x_1)\phi(x_2)\phi^*(x_3)\phi(x_4)G_1(x_1-x_2)\times\nn\\
&\quad& \times G_2(x_2-x_3)G_1(x_3-x_4)G_2(x_4-x_1).
\ee
As for $\delta b_2,$ it is convenient to work in momentum space. Equation (\ref{32}) now receives extra contributions coming also from these higher terms in the effective action, as well as those coming from higher orders from $D^{-1}(\omega,k)$. Collecting all the similar terms one then systematically finds $\delta \Omega_3$, $\delta \Omega_4$, ..., and one then uses Eq. \eqref{28} to find $\delta b_3,\delta b_4$, ... The actual calculations will require explicit treatment of Matsubara sums  (just as for $\delta b_2$).\\

\subsection{Sketch of the calculation of ${\delta b_3}$}

\noindent While the calculational scheme for $\delta b_n$ described above is systematic and straightforward, the actual details are not trivial. We will present here the first details, including preliminary numerical evaluations, of $\delta b_3$. Beyond what will be discussed below, we have produced further analytical expressions coming from the Wick expansion term, Eq. \eqref{wick} below. Extensive numerical work is currently underway; full details will be published elsewhere.\\

\noindent Let us start by writing the quadratic part of the grand potential, Eq. (\ref{32}), as
\small
\be 
\delta\Omega_0 = \frac{1}{2\pi i}\int_{-\infty}^{\infty}d\omega\int\frac{d^2k}{(2\pi)^2}\text{Disc}\left\{\ln\mathcal{D}^{-1}(\omega+i\epsilon,k)\right\}f_\text{BE}(\omega),\nn\\
\label{om3}
\ee
\normalsize
which comes from the quadratic partition function
\be 
\Z_0=\int\frac{[d\phi^* d\phi]}{N}e^{-S_2}.
\ee
On the other hand, the general form for the partition function is written as
\be 
\Z&=&\Z_\text{free}\int\frac{[d\phi^* d\phi]}{N}e^{-(S_2+\delta S)}\nn\\
&=&\Z_\text{free}\Z_0\left(1-\Z_0^{-1}\int\frac{[d\phi^* d\phi]}{N}e^{-S_2}\delta S\right)\nn\\
&=&\Z_\text{free}e^{-\beta\delta \Omega},\quad \delta\Omega=\delta\Omega_0+\delta \tilde{\Omega}.
\ee
$\delta b_3$ is therefore expected to have contributions from both $\delta\Omega_0$ and $\delta\tilde{\Omega}$.
\begin{itemize}
\item From $\delta\Omega_0$: Let us remember that the term $\D^{-1}$ has an expansion in the fugacity\footnote{This comes when one expands the Fermi Dirac distribution as $f_k=ze^{-\beta k^2/2}-z^2e^{-\beta k^2}+O(z^3)$ in $f_k=\frac{1}{e^{\beta (k^2/2-\mu)}+1}$.} $z$
\be
\D^{-1}
&=&\D^{-1}(z^0)\left(1-z\mathcal{D}(z^0)B\right)+O(z^2),
\ee
where
\be
B  = \int \frac{d^2k}{(2\pi)^2}\frac{e^{-\beta k^2/2}+e^{-\beta (k+q)^2/2}}{\omega+i\epsilon-\left(\frac{k^2}{2}+\frac{(k+q)^2}{2}\right)+2\mu},
\ee
and hence
\be \label{equation39}
\ln\D^{-1} 
&=&\ln\D^{-1}\left(z^0\right)-z\mathcal{D}(z^0)B+O(z^2).
\ee
%
\item From $\delta\tilde{\Omega}$\footnote{Note $\langle\hat{A}\rangle_0=\Z_0^{-1}\int \frac{\left[d\phi^*d\phi\right]}{N}e^{-s_2}A.$}:
\be  \label{equationC}
\Z&=&\Z_\text{free}\Z_0\left[1-\Z_0^{-1}\int\frac{[d\phi^* d\phi]}{N}e^{-S_2}\delta S\right]\nn\\
&=&\Z_\text{free}\Z_0\left[1-\langle\delta S\rangle_0\right]=\Z_\text{free}\Z_0\left[1-C\right].
\label{ceq}
\ee
Using Eq. \eqref{equation33}, and defining $x_{ij} = x_i - x_j$, the quantity $C$ is given by 
\be
\label{wick}
 C&=&\frac{1}{2}\int \prod_{i=1}^4dx_iG_1(x_{12})G_2(x_{23})G_1(x_{34})G_2(x_{41})\times\nn\\
&\quad& \times \langle\phi^\dag(x_1)\phi(x_2)\phi^\dag(x_3)\phi(x_4)\rangle_0,
\ee
where what is left to do is evaluate the expectation value by using Wick's theorem, taking into account the order in fugacity for the product in the $G$s. This is to be done in momentum space ($x\to (\omega_n,\vec{k})$) such that
\be 
 \contraction{}{\phi}{^*(y)}{\phi}\phi^*(y)\phi(x) \equiv \Delta(y-x)
 \label{delt}
\ee
will introduce terms proportional to the Bose-Einstein distribution and therefore both Fermi and Bose Matsubara sums will appear. 
Using Eq.~\eqref{equation39} in Eq.~\eqref{om3}, one can show that the contribution from $\delta \Omega_0$ to $\delta b_3$ is
\begin{equation}\label{equation44}
 \delta b_3^0=4\int^\infty_0\frac{d\omega \,e^{-4\omega}}{\ln\left(\frac{\beta E_b}{3\omega}\right)^2+\pi^2}\left(E_i(\omega)+\ln\left(\frac{\beta E_b}{3\omega}\right)\right),
\end{equation}
where $E_i(x)$ is the exponential integral \cite{1965hmfw}. 
\item Fig. 2 shows $\delta b_3^0$ vs $\beta E_b$. The comparison with the results by the authors of ref. \cite{Parish} who computed $\delta b_3$ using other methods shows that it is indeed necessary to compute the Wick terms contribution to $\delta b_3$, the analytical expressions of which we have. Once the numerical evaluation is completed we will compare with Ref.~\cite{Parish} in a forthcoming publication \cite{unpublished}.
\begin{figure}[t]
		\begin{center}
		\includegraphics[scale=0.7]{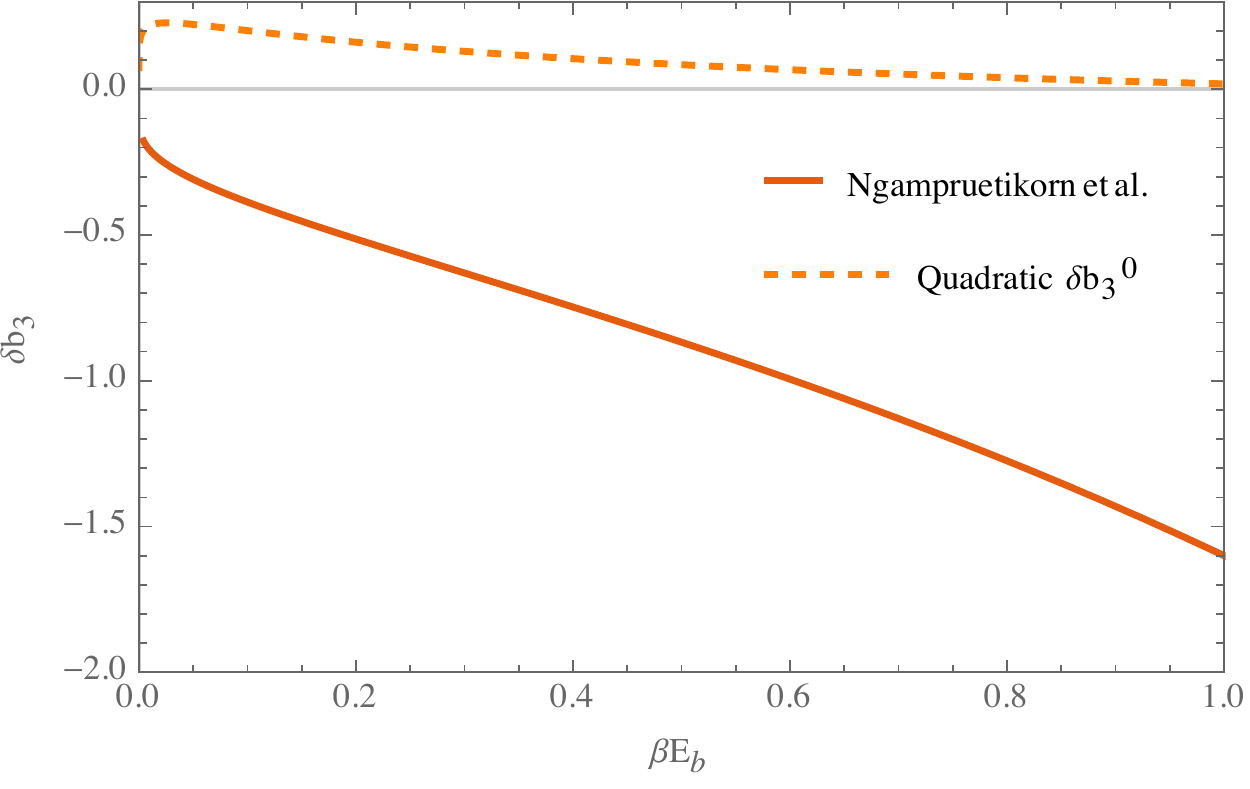}
		\caption{Contribution from $\delta \Omega_0$ to $\delta b_3$ compared with \cite{Parish}.}
		\label{figure2}
	\end{center}
	\end{figure}

\end{itemize}

\section{Conclusions and Comments}

\noindent In this paper we have demonstrated an intimate connection between 2D $SO(2,1)$ scaling anomalies and the existence of the Tan contact, namely, the contact is essentially the anomaly, Eq. \eqref{equation11}.  This identification allowed us to derive an expression for the shift of the n-th virial coefficient, Eq. \eqref{28}, in terms of $\frac{\partial \left(\delta \Omega_n\right)}{\partial E_b}$, where $\delta \Omega_n$ is the corresponding part of $\Omega$ coming from interactions, Eq. \eqref{Omega}. In particular, we were able to derive $\delta b_2$, which coincides with the Beth-Uhlenbeck formula, validating in this fashion the original heuristic motivation for this work, i.e., the connection between 2D $SO(2,1)$ scaling anomalies and the non-zero value of $\delta b_2$. In the process, we also derived the n-th virial expansion for the Tan contact, and a systematic and self-consistent procedure to calculate $\delta b_n$, $n \geq 3$ was developed through a formal expansion of the path integral. Partial results for $\delta b_3$ were discussed; the full calculation will be reported elsewhere.  We have also recently discovered a mapping of the anomalous 2D two-body contact interaction studied in this paper and the anomalous 1D three-body contact interaction \cite{Drut:2018rip}.  Applications of these ideas to other systems with $SO(2,1)$ symmetry  in molecular, atomic, condensed-matter, high-energy and biological physics are underway. \\

\acknowledgments
We are most thankful to J. Levinsen and M. Parish for illuminating conversations and for sharing their numerical values of $\delta b_3$ with us. We would also like to thank T. Schaefer for his helpful comments and advice, as well as P. Hosur for comments on our work. This work was supported in part by the US Army Research Office Grant No. W911NF-15-1-0445 and the U.S. National Science Foundation under Grant No. PHY1452635 (Computational Physics Program).


%


\appendix
\section{Derivation of Eq. \eqref{6equation}}

Consider the set of microscopic parameters $g_j$ (coupling constants from the Lagrangian). We can form energy parameters $E_j$ taking suitable powers of $g_j$; consider also possible bound states and energies of the system, $E_{b\ell}$, hence forming the set of energy parameters $E_k=\{E_j,E_{b\ell} \}$. The grand thermodynamical potential $\Omega=\Omega(\beta, \mu_i, V,E_k)$ for a homogeneous system in D-spatial dimensions must have the form ($\Omega$ is an extensive variable)

\begin{equation}
\Omega(\beta, \mu_i, V,E_k)=V\beta^{-1-D/2}f(z_i,\beta E_k),
\end{equation} 

\noindent where $f(z_i,\beta E_k)$ is a dimensionless function of dimensionless variables and $z_i=e^{\beta \mu_i}$ is the fugacity corresponding to $\mu_i$. It is straightforward to show that \cite{Lin:2016shf}

\begin{equation}
\beta \frac{\partial \Omega}{\partial \beta} \Bigg |_{z_i,V}=\left(-1-\frac{D}{2}\right)\Omega+\sum\limits_k E_k \frac{\partial \Omega}{\partial E_k}.
\end{equation}

\noindent Using the thermodynamic identity $E=\frac{\partial(\beta \Omega)}{\partial \beta} \big |_{z_i,V}=\Omega+\beta \frac{\partial \Omega}{\partial \beta}\big |_{z_i,V}$ we get (also use $\Omega=-PV$)

\begin{align}
2E-DPV&=2\left(\Omega+\beta \frac{\partial \Omega}{\partial \beta}\Bigg |_{z_i,V}\right)-DPV \nn \\
&=2\left(\Omega-\left(1+\frac{D}{2}\right)\Omega+\sum\limits_k E_k\frac{\partial \Omega}{\partial E_k} \right)-DPV \nn\\
&=-2\sum\limits_k E_k\frac{\partial P}{\partial E_k}V. 
\end{align}

\noindent Therefore

\begin{equation}
2\mathcal E-DP=-2\sum\limits_k E_k\frac{\partial P}{\partial E_k}.
\end{equation}

\section{Heuristic proof of Eq. \eqref{equation11} }

\noindent Consider the partition function for the scaled sytem $\tau \rightarrow \lambda^2 \tau$, $\vec{x}\rightarrow \lambda \vec{x}$ 

\begin{align}
Z \rightarrow Z^\lambda&=\int \left[ d\psi_\sigma^{*\lambda} d\psi_\sigma^{\lambda}  \right]\,e^{-S^{\lambda}[\psi^*,\psi]} \nonumber\\
&=J\tilde{Z},
\end{align}
where
\be
S^\lambda[\psi^*,\psi] \equiv
\int^\beta_0 \!\! d\tau \!\! \int \! d^2\vec{x} \,\left[
\psi_\sigma^{*}\left(\partial_\tau\!-\!\frac{\nabla^2}{2}\!-\!\tilde{\mu}\right)\psi_\sigma
+c^\lambda \psi^*_\uparrow \psi^*_\downarrow \psi_\downarrow \psi_\uparrow
\right]\nn\\
\ee
and
where $J$ is the Jacobian of the transformation $\psi_\sigma$, $\psi^*_\sigma$ $\rightarrow$ $\psi^\lambda_\sigma$, $\psi^{*\lambda}_\sigma$, $\tilde{\mu}=\lambda^2 \mu$, and $\tilde{Z}$ is
\begin{equation}
\tilde{Z}=\text{tr}\,\left(e^{-\beta \left( H(\lambda) -N\tilde{\mu}\right)} \right),
\end{equation}
where
\begin{equation}\label{equation47}
H(\lambda)=\int d^2\vec{x} \, \left(H_0+c^\lambda \psi^*_\uparrow \psi^*_\downarrow \psi_\downarrow \psi_\uparrow \right).
\end{equation}

In Eq. \eqref{equation47} $H_0$ is the free Hamiltonian and $c^\lambda$ is the rescaled coupling constant (under $\vec{k}\rightarrow \lambda^{-1}\vec{k}$) 
\begin{align}
\frac{1}{c} \rightarrow \frac{1}{c^\lambda}&=\frac{1}{(2\pi)^2}\int^{\tilde{\Lambda}=\lambda^{-1}\Lambda \rightarrow \infty} \!\!\!\!\!\!\frac{d^2\vec{\tilde{k}}}{-E_b-\vec{\tilde{k}}^2+i\epsilon} \nonumber \\
&=\frac{1}{(2\pi)^2}\int^{\Lambda \rightarrow \infty}\!\!\!\!\!\!\frac{d^2\vec{k}}{-\lambda^2E_b-\vec{k}^2+i\epsilon}=\frac{1}{c\left(\frac{\lambda^2 E_b}{\Lambda^2}\right)}.
\end{align}
Under an infinitesimal dilation $\lambda=1+\delta \lambda$,
\be
\delta Z\big |_{\lambda=1} &\equiv& Z^{\lambda=1+\delta \lambda}-Z \nn \\
&=&\delta J(\lambda)\tilde{Z}\big |_{\lambda=1}+J(\lambda)\big |_{\lambda=1}\left(\frac{\partial \tilde{Z} }{\partial \lambda}\right)\bigg |_{\lambda=1}\delta \lambda.\quad
\ee
It is straightforward to show that
\begin{equation}\label{equationb6}
\delta Z\big |_{\lambda=1}=\delta J(\lambda)\big |_{\lambda=1}Z+2Z\left[
\mu \beta \langle N \rangle+\int^\beta_0 d\tau \, \langle I \rangle
\right]\delta \lambda,
\end{equation}
where the angle brackets $\langle \, \, \rangle$ denote the thermal average and $I$ is Tan's contact
\begin{equation}
I=\frac{c^2}{4\pi} \int d^2\vec{x}\, \psi^\dagger_\uparrow \psi^\dagger_\downarrow \psi_\downarrow \psi_\uparrow.
\end{equation}
On the other hand, in the large $A$ (volume in 2D) limit,
\begin{equation}
Z=e^{-\beta \Omega}=e^{\beta P A},
\end{equation}
and
\begin{equation}
Z^\lambda=e^{\beta^\lambda P^\lambda A^\lambda},
\end{equation}
and with $\lambda=1+\delta \lambda$, using thermodynamic identities~\cite{Ordonez:2015vaa}, after some algebra one obtains
\begin{equation}\label{equationb9}
\delta Z=2\beta Z\left[\mu \langle N \rangle +PA-\langle H\rangle 
\right]\delta \lambda.
\end{equation}
Comparing Eqs. \eqref{equationb6} and \eqref{equationb9} we get ($\mathcal{E}=\frac{\langle H \rangle}{A}$)

\begin{equation}
PA-\langle H \rangle=\text{Jacobian term}+\langle I \rangle.
\end{equation}
In references \citep{Ordonez:2015vaa,LinCarlos,pathintegral1,pathintegral2,pathintegral4,Lin:2016shf} the Jacobian term was shown to be proportional to $c^2\left(\psi^*_\uparrow \psi_\downarrow \right)^2$ where $\psi^*_\uparrow \psi_\downarrow $ is a constant background value (finite). In our case, $c= c(E_b/\Lambda^2) \rightarrow 0$ when $\Lambda\rightarrow \infty$, and below, when we calculate the virial coefficients, an expansion around $\psi^*_\uparrow \psi_\downarrow =0$ will be performed. In either case, the Jacobian term in this case is zero and the anomaly is completely captured by the Tan contact \cite{Hofmann:2012np}. The final result is then 

\begin{equation}
\text{Anomaly}=\mathcal{ A}=2P-2\mathcal{E}=\frac{2}{A}\langle I \rangle.
\end{equation}

\section{Derivation of Eq. (\ref{deriv})}
\noindent The definition for the complex logarithm is
\be 
\ln(x+iy)=\ln\sqrt{x^2+y^2}+i\text{Arg}(y,x),
\ee
where

\be 
\text{Arg}(y,x)=\left\{\begin{matrix}
\arctan(y/x),&x>0\\
\arctan(y/x)+\pi,&x<0, y\geq0\\
\arctan(y/x)-\pi,&x<0, y<0\\
\pi/2,&x=0,y>0\\
-\pi/2,&x=0,y<0\\
\text{Undefined},& x=y=0.
\end{matrix}
\right.
\ee
Let us analyze the different possibilities for  $\omega$ in $h\equiv \text{Disc} (\ln\D^{-1})$:
\be
h&=&\ln\left[\frac{1}{4\pi}\ln\left(-\frac{\omega+2\mu-\frac{\varepsilon_k}{2}}{E_b}-i\epsilon\right)\right]\\
&&-\ln\left[\frac{1}{4\pi}\ln\left(-\frac{\omega+2\mu-\frac{\varepsilon_k}{2}}{E_b}+i\epsilon\right)\right].
\ee
Therefore we recognize two regions for the variable $\omega$
\begin{itemize}
\item $\omega<\frac{\varepsilon_k}{2}-2\mu$, where
\be
h
&=&\left\{\begin{matrix}
0,&\omega<\frac{\varepsilon_k}{2}-2\mu-E_B,\\
-2i\pi\equiv h_1,& \frac{\varepsilon_k}{2}-2\mu-E_B<\omega<\frac{\varepsilon_k}{2}-2\mu.\\
\end{matrix}\right.
\ee
\item $\omega>\frac{\varepsilon_k}{2}-2\mu$, where
\be 
h
&=& \left\{\begin{matrix}
-2\pi i + h_3 \equiv h_2,&\frac{\varepsilon_k}{2}-2\mu<\omega<\frac{\varepsilon_k}{2}-2\mu+ E_B,\\
h_3,&\omega>\frac{\varepsilon_k}{2}-2\mu+E_B.\\
\end{matrix}\right.\nn\\
\ee
\end{itemize}
where
\be
h_3 \equiv -2i\arctan\left[\frac{\pi}{\ln{\left(\frac{\omega+2\mu-{(1/2)\varepsilon_k}}{E_B}\right)}}\right].
\ee

The results for the regions of $\omega$ are summarized in Table (\ref{table1}).\\

\begin{table}
\caption{Discontinuities and drops. }\label{table1}
\centering
\begin{tabular}{|c|c|c|}
\hline \hline
$\omega$&Discontinuity $(h_i)$&Drop\\
\hline
$(\infty,\varepsilon_k/2-2\mu-E_B)$ &0 & --- \\
$(\varepsilon_k/2-2\mu-E_B,\varepsilon_k/2-2\mu)$&$-2\pi i$&$+2\pi i$\\
$(\varepsilon_k/2-2\mu,\varepsilon_k/2-2\mu+E_B)$&$-2\pi i + h_3 $&0\\
$(\varepsilon_k/2-2\mu+E_B,\infty)$&$h_3$&$0$\\
\hline
\end{tabular}
\end{table}
\noindent Consider the following expression
\be 
\int^{\infty}_{a(t)}h(x,t)dx&=&\int^{b(t)}_{a(t)}h_1(x,t)dx+\int_{b(t)}^{c(t)}h_2(x,t)dx \nn  \\ 
&& +\int_{c(t)}^{\infty}h_3(x,t)dx,
\ee
where $h=\text{Disc}(\ln\mathcal{D}^{-1}),$ $x=\omega$ and $t=E_B$. Following Eq.(\ref{28}) we need to take 
the derivative with respect to $t$
\begin{widetext}
\be 
\frac{\partial}{\partial t}\int^{\infty}_{a(t)}h(x,t)dx&=&
-\frac{\partial a(t)}{\partial t}h_1(a(t),t)+\frac{\partial b(t)}{\partial t}\left[h_1(b(t),t)-h_2(b(t),t)\right]+\frac{\partial c(t)}{\partial t}\left[h_2(c(t),t)-h_3(c(t),t)\right]\nn\\
&&\quad+\int^{b(t)}_{a(t)}\frac{\partial}{\partial t}h_1(x,t)dx+\int_{b(t)}^{c(t)}\frac{\partial}{\partial t}h_2(x,t)dx+\int_{c(t)}^{\infty}\frac{\partial}{\partial t}h_3(x,t)dx \nn\\
&&=-\frac{\partial a(t)}{\partial t}h_1(a(t),t)+\frac{\partial b(t)}{\partial t}\left[Drop_1\right]+\frac{\partial c(t)}{\partial t}\left[Drop_2\right]+\int^{b(t)}_{a(t)}\frac{\partial}{\partial t}h_1(x,t)dx\nn\\
&&\quad+\int_{b(t)}^{c(t)}\frac{\partial}{\partial t}h_2(x,t)dx+\int_{c(t)}^{\infty}\frac{\partial}{\partial t}h_3(x,t)dx.
\ee
\end{widetext}

Here $Drop_i$ corresponds to the drop of the function $h_i$ when it argument $x$ goes from $x-\delta$ to a value $x+\delta$ with $\delta<<1.$  These terms are recorded in Table {\ref{table1}}.




\noindent We obtain the following expression
\begin{widetext}
\begin{eqnarray}
\frac{\partial \delta\Omega}{\partial E_B}&=&
\frac{1}{2\pi i}\int\frac{d^2k}{(2\pi)^2}\left\{\frac{\partial}{\partial E_B}(\frac{\varepsilon_k}{2}-2\mu-E_B)(2i\pi)f_B(\omega=\xi_k)+\int_{\varepsilon_k/2-2\mu}^{\infty}\frac{\partial h_3}{\partial E_B} f_B(\omega)d\omega\right\}\nonumber\\
&&\!\!\!\!\!\!\!\!\!\!=\frac{1}{2\pi i}\int\frac{d^2k}{(2\pi)^2}\Bigg\{-z^2e^{\beta E_B}(2\pi i)e^{-\beta \varepsilon_k/2}\left.-\int_{\varepsilon_k/2-2\mu}^{\infty}d\omega\int\frac{d^2k}{(2\pi)^2}\frac{f_B(\omega)(2i\pi)}{E_B\left(\pi^2+\ln^2\left(\frac{\omega+2\mu-\frac{\varepsilon_k}{2}}{E_B}\right)\right)}\right\}+O(z^3)\nn\\
&&\!\!\!\!\!\!\!\!\!\!=-\frac{z^2}{\pi\beta}e^{\beta E_B} -z^2\int_0^\infty d\tilde{\omega} \int\frac{d^2k}{(2\pi)^2} \frac{e^{\beta(\tilde{\omega}-\frac{\epsilon_k}{2})}}{E_B\left(\pi^2+\ln^2(\frac{\tilde{\omega}}{E_B})\right)}+O(z^3)\nn\\
&&\!\!\!\!\!\!\!\!\!\!=-\frac{z^2}{\pi\beta}\left(e^{\beta E_B} +2\int_0^\infty d\tilde{k} \tilde{k}\frac{e^{-\beta\tilde{k}^2}}{E_B\left(\pi^2+\ln^2(\frac{\tilde{k}^2}{E_B})\right)}\right)+O(z^3),
\end{eqnarray}
\end{widetext}

\noindent where $\xi_k \equiv \varepsilon_k/2-2\mu-E_B$, we have used the change of variables $\omega=\tilde{\omega}-2\mu+ \frac{\epsilon_k}{2}$, and the substitution 
$\tilde{\omega}\to \tilde{k}^2$. 

\noindent Thus, at second order in the fugacity we obtain
\be  
\frac{\partial\delta \Omega_2}{\partial E_b}=-\frac{1}{\pi\beta}\left(e^{\beta E_b} +2\int_0^\infty d\tilde{k} \tilde{k}\frac{e^{-\beta\tilde{k}^2}}{E_b\left(\pi^2+\ln^2(\frac{\tilde{k}^2}{E_b})\right)}\right).\nn\\
\ee

\end{document}